\documentclass[
 reprint,
 superscriptaddress,
 amsmath, amssymb,
 prb,
 citeautoscript, 
]{revtex4-1}

\usepackage[utf8]{inputenc}
\usepackage{graphicx}
\usepackage{dcolumn}
\usepackage{bm}
\usepackage{tikz}
\usetikzlibrary{shapes.geometric, arrows.meta, positioning, calc}
\usepackage[utf8]{inputenc}
\usepackage{listings}

\usepackage{xcolor}
\usepackage{soul}

\newcommand{\MS}[1]{\textcolor{red}{#1}}

\definecolor{codegreen}{rgb}{0,0.6,0}
\definecolor{codegray}{rgb}{0.5,0.5,0.5}
\definecolor{codepurple}{rgb}{0.58,0,0.82}
\definecolor{backcolour}{rgb}{0.96,0.96,0.98}

\lstdefinestyle{paperstyle}{
    backgroundcolor=\color{backcolour},   
    commentstyle=\color{codegray},
    keywordstyle=\color{magenta},
    numberstyle=\tiny\color{codegray},
    stringstyle=\color{codepurple},
    basicstyle=\ttfamily\small,
    breakatwhitespace=false,         
    breaklines=true,                 
    captionpos=b,                    
    keepspaces=true,                 
    numbers=left,                    
    numbersep=8pt,                  
    showspaces=false,                
    showstringspaces=false,
    showtabs=false,                  
    tabsize=4,
    frame=single,
    rulecolor=\color{codegray!30},
    language=Python
}

\definecolor{mainblue}{RGB}{35, 85, 125}
\definecolor{softgray}{gray}{0.98}
\definecolor{linegray}{gray}{0.4}

\lstdefinestyle{compactpython}{
    language=Python,
    basicstyle=\ttfamily\footnotesize, 
    backgroundcolor=\color{softgray},
    keywordstyle=\color{mainblue}\bfseries,
    commentstyle=\color{gray}\itshape,
    stringstyle=\color{green!40!black},
    numbers=left,
    numberstyle=\tiny\color{linegray},
    stepnumber=1,
    numbersep=5pt,
    frame=l,                  
    framerule=1pt,
    rulecolor=\color{mainblue!30},
    xleftmargin=15pt,
    showstringspaces=false,
    breaklines=true,
    lineskip=-1pt,            
    aboveskip=10pt,
    belowskip=10pt,
    escapeinside={(*@}{@*)}   
}

\begin{document}

\preprint{APS/123-QED}

\title{VASP Plugins: Linking the Vienna \textit{ab-initio} Simulation Package with Python}

\author{Sudarshan Vijay}
\thanks{Equal contributions}
\email{sudarshan.vijay@iitb.ac.in}
\affiliation{Department of Chemical Engineering, Indian Institute of Technology Bombay, Powai, Mumbai, Maharashtra 400076 India}
\author{Martijn Marsman}
\affiliation{VASP Software GmbH, Berggasse 21, 1090 Vienna, Austria}
\author{Georg Kresse}
\email{georg.kresse@univie.ac.at}
\affiliation{VASP Software GmbH, Berggasse 21, 1090 Vienna, Austria}
\affiliation{Faculty of Physics and Center for Computational Materials Science, University of Vienna, Kolingasse 14-16, A-1090 Vienna, Austria}%
\author{Martin Schlipf}
\thanks{Equal contributions}
\email{martin.schlipf@vasp.at}
\affiliation{VASP Software GmbH, Berggasse 21, 1090 Vienna, Austria}

\date{\today}

\begin{abstract}
Implementing novel features and experimental algorithms into widely adopted density functional theory (DFT) codes is frequently hindered by complex legacy architectures and the use of compiled languages such as Fortran. These production codes, while optimised for high-performance computing clusters, present significant hurdles for software development and rapid prototyping, often requiring deep expertise in the code's internal structure to modify. To address this challenge, we present a Python plugin infrastructure for the Vienna \textit{ab-initio} Simulation Package (VASP) that combines computational efficiency with the flexibility of high-level scripting. Our architecture uses a C++ intermediate layer and \verb|pybind11| to expose VASP data as \verb|NumPy| arrays via shared memory buffers, ensuring high performance without data duplication. We implement two categories of plugins: those that modify quantities at the end of each converged self-consistent field (SCF) cycle, such as \verb|structure| and \verb|force_and_stress|, and those that operate during the SCF cycle, such as \verb|local_potential| and \verb|occupancies|. We  demonstrate the utility of our implementation through three applications, structure relaxation using the \verb|scipy| library, implementing an implicit solvent model, and adding the \verb|DFT-D4| dispersion corrections. This infrastructure effectively bridges the gap between high-performance electronic structure routines and the widespread scientific Python ecosystem.
\end{abstract}

\maketitle

\section{Introduction}
\label{sec:introduction}

Materials discovery and property prediction rely on accurate and rapid first principles computational methods such as density functional theory (DFT).\cite{eckertAFLOWLibraryCrystallographic2024,jainFormationEnthalpiesMixing2011}
These calculations are often performed on high-performance computing clusters.\cite{Kresse1996,zhaoVASPPerformanceHPE}
Given this focus on performance, most production DFT codes are written in compiled languages such Fortran\cite{Kresse1996,Giannozzi2009} and often focus on improving computational speed and numerical stability of their underlying codebase.

These large-scale Fortran projects present significant hurdles for modern software development and rapid prototyping.
Programs such as the Vienna \textit{ab-initio} Simulation Package (VASP),\cite{Kresse1996} a widely used DFT code,\cite{talirzTrendsAtomisticSimulation2021} are often characterized by complex memory management and architectures that make it difficult for researchers to implement specific features or experimental algorithms.
Consequently, the barrier to entry for modifying the source code is high, frequently requiring deep expertise in the legacy structure of the program.

To circumvent the need for direct source code modifications, researchers frequently rely on Python-based frameworks such as the Atomic Simulation Environment (ASE),\cite{Larsen2011} AiiDA,\cite{Huber2020,Huber2021} and Phonopy\cite{togoFirstPrinciplesPhonon2015}.
These packages operate as wrappers, automating complex computational workflows by generating input files, executing the compiled binaries, and parsing the resulting output files.
While these tools are highly effective for high-throughput screening and post-processing, their reliance on file input/output (I/O) restricts them to an external, black-box operation of the DFT engine.
Due to this black-box nature of their implementation, they cannot interact dynamically with the computations during runtime.

To implement an interface with more direct control, there is a growing trend toward hybrid software architectures that combine the computational efficiency of Fortran with the flexibility of high-level scripting languages.\cite{VanDerWalt2011,Virtanen,mortensenGPAWOpenPython2024}
Python has become the language of choice for the computational physics community, offering an extensive ecosystem of libraries for data analysis, plotting, numerical methods, and machine learning.
By implementing an interface  between the VASP and Python, it is possible to maintain the speed of the optimized electronic structure routines while providing ready customisation.

In this work, we present the Python plugin infrastructure for VASP.
This implementation allows users to write Python functions to modify the behaviour of VASP at specific locations in the source code.
We implement two sets of plugins.
The first set modifies or replaces selected quantities at the end of each converged self-consistent field (SCF) cycle.
For example, the \verb|structure| plugin modifies the coordinates and lattice vectors and the \verb|force_and_stress| plugin modifies or replaces the forces and the stress tensor.
The second set of plugins modifies selected quantities during the SCF cycle.
For example, the \verb|local_potential| plugin modifies the total electrostatic potential and the \verb|occupancies| plugin modifies the number of electrons, the Fermi energy, and smearing parameters.
The manuscript is structured as follows: in Section \ref{sec:architecture}, we describe the VASP Python plugin infrastructure. In Section \ref{sec:illustrations} we illustrate its utility with a few common use-cases in electronic structure calculations.
\section{Architecture}
\label{sec:architecture}

\begin{figure*}[!htb]
    \centering
    \includegraphics[width=0.7\linewidth]{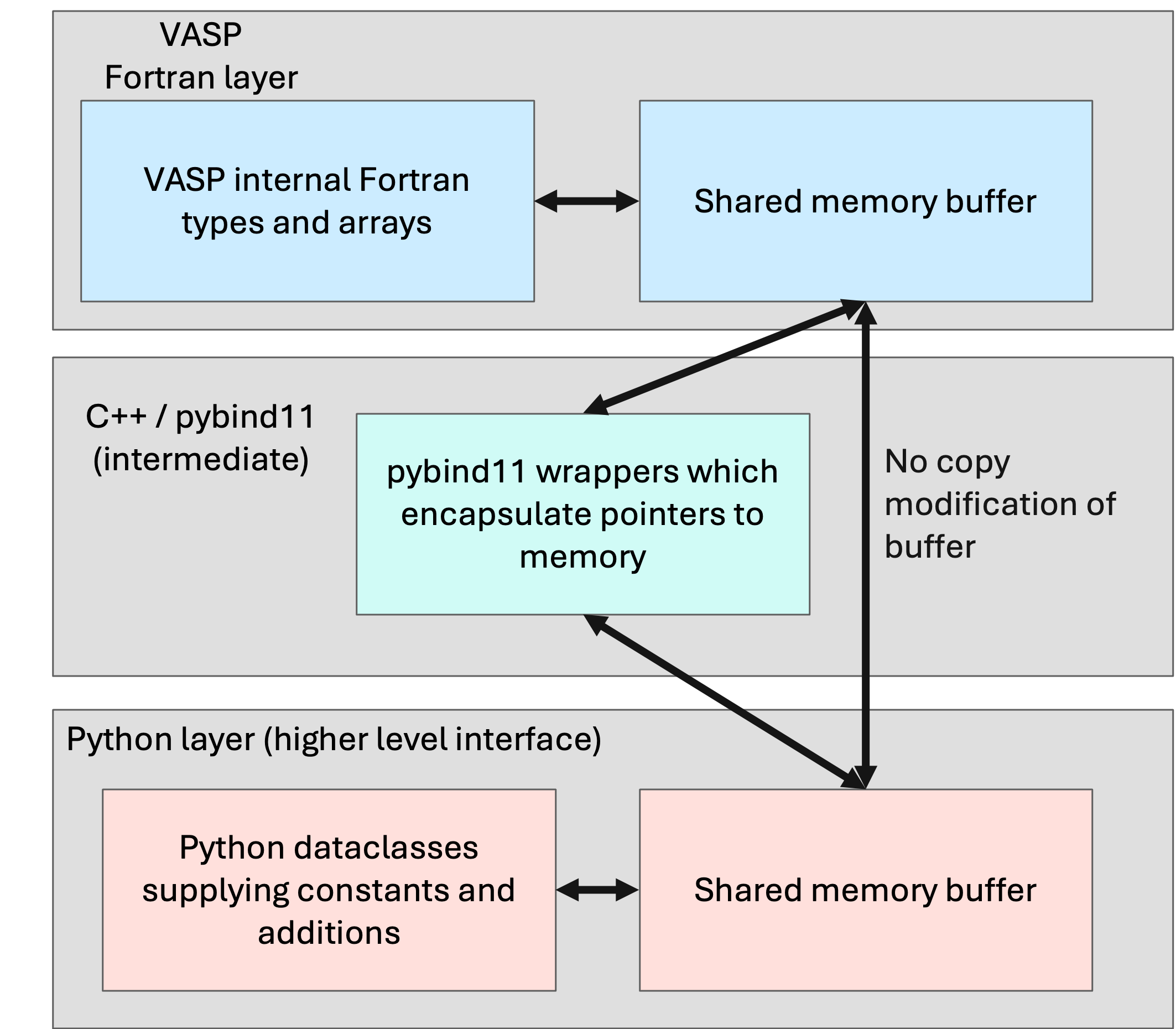}
    \caption{Architecture of the Fortran-Python interface. Data is loaded into a buffer in the Fortran layer and exposed to Python via pybind11 as NumPy arrays. Python modifies the shared memory directly, ensuring high performance without data duplication.}
    \label{fig:architecture}
\end{figure*}

Figure \ref{fig:architecture} shows a schematic of the architecture of VASP Python plugins.
The top layer of this architecture consists of the VASP Fortran code.
All integer, real\MS{,} and complex arrays are created in this layer. 
This layer operates as a regular VASP calculation, with no change until it reaches critical ``hooks'' in the source code.

A shared memory buffer is created once these hooks are reached.
This shared memory buffer is composed entirely of data types compatible with C using the \verb|ISO_C_BIND| module.
These C-bind attributes prepare the data to be transferred from Fortran to Python through C++ and back.
Pointers are stored instead of a copy of the data for large grid quantities (such as the charge density). 
This strategy prevents any memory overhead when transferring data between Fortran and Python.

We separate data into two classes, depending on their intended use.
Data that should not be modified through the interface are stored in the \verb|constants| data structure and data that is expected to be modified are stored in the \verb|additions| data structure.
For example, in the \verb|structure| plugin, the \verb|constants| data structure consists of unmodifiable input quantities such as number of ions, ion types and unmodifiable output quantities such as the total energies and forces. The \verb|additions| data structure consists of modifiable quantities such as lattice vectors and the atom positions.

A C++ layer is used as an intermediate to transfer \verb|constants| and \verb|additions| data structures between Fortran and Python.
Data is transferred between Fortran and C++ using a Fortran interface function.
All interface functions take only \verb|constants| and \verb|additions| data structures as inputs.

The \verb|constants| and \verb|additions| data structures are converted to \verb|numpy| arrays using the \verb|pybind11| package.\cite{PybindPybind112026}
Three dimensional grids, such as the charge density, are reshaped to the appropriate dimensions and converted to \verb|numpy| arrays.
All \verb|numpy| arrays are stored in Python dataclasses with the same arguments.
These dataclasses hold all \verb|constants| and \verb|additions| data structures, analogous to their Fortran and C++ counterparts.

\begin{lstlisting}[language=Python, caption={Constants for the structure plugin; the frozen keyword argument is added to prevent inadvertent modification of constants.}, label={lst:structure}]
@dataclass(frozen=True)
class ConstantsStructure:
    number_ions: int
    number_ion_types: int
    ion_types: IndexArray
    atomic_numbers: IntArray
    lattice_vectors: DoubleArray
    positions: DoubleArray
    POMASS: DoubleArray
    total_energy: float
    forces: DoubleArray
    stress: DoubleArray
    shape_grid: IntArray
    charge_density: DoubleArray
    neighbor_list: List[Neighbors]
\end{lstlisting}

Analogous to the Fortran and C++ interface functions, the Python functions also take \verb|constants| and \verb|additions| data structures as input.
For example, the \verb|constants| for the \verb|structure| plugin (see Listing \ref{lst:structure}) shows the type and data structures which are accessible to the \verb|structure| Python function.

To use this plugin infrastructure, users simply write a Python function in a file called \verb|vasp_plugin.py| with a predefined name that takes in \verb|constants| and \verb|additions|.
For instance, to modify the structure, we need to specify a function\MS{:} \verb|structure(constants, additions)|.
Any modification to the \verb|additions| dataclass must be done through the \verb|+=| or \verb|-=| operators.

Note that the \verb|vasp_plugin.py| file is treated as a regular Python file to be processed by the Python interpreter.
This file offers all the flexibility associated with programming in Python, with the only constraint being that the pre-determined functions take \verb|constants| and \verb|additions| as inputs.
\section{Illustrations}
\label{sec:illustrations}

In this section, we discuss three applications of our VASP Plugins implementation.
First, we discuss the combination of the VASP SCF procedure with external structure relaxation tools using the \verb|structure| plugin.
As an illustration, we relax the atom-centred forces using implementations from the \verb|scipy| Python package.\cite{Virtanen}
Second, we implement an implicit solvent package entirely in Python and link it with the \verb|local_potential| plugin interface.
Finally, we discuss how the atom-centred forces can be modified using the \verb|force_and_stress| plugin.
We illustrate this modification by adding dispersion corrections through the \verb|DFT-D4|\cite{caldeweyherGenerallyApplicableAtomiccharge2019,caldeweyherExtensionEvaluationD42020} package.

\subsection{Relaxation using the \texttt{scipy} library and the \texttt{structure} plugin}

In this illustration, we use two  optimisers implemented within the \verb|scipy| Python package to relax the forces on a structure.
We implement a link between two optimisers implemented in the \verb|scipy.minimize| function, conjugate gradient (CG) and the Broyden-Fletcher-Goldfarb-Shanno (BFGS).

Listing \ref{lst:illustration_scipy} shows the code to be added to the \verb|vasp_plugin.py| file to link \verb|scipy| as a back-end for structure relaxation.
The \verb|history| dictionary stores a tuple of the energies and its gradient.
In the \verb|structure| function, positions and lattice vectors at the end of each SCF cycle is accessed through the \verb|constants| dataclass.
Note that the positions are sent as fractional coordinates and would need to be converted to cartesian coordinates to be commensurate with the forces during relaxation.
The initial structure is stored in  \verb|x_start|, which will eventually be passed to the \verb|minimize function|.
Two functions, \verb|fun| and \verb|jac| return the total energy and its gradient respectively, if it is stored in history.

The minimize function is called with \verb|x_init|, \verb|fun|, \verb|jac| as arguments.
Each call to \verb|structure| restarts this \verb|minimize| function until it finds a structure that does not exist in \verb|history|.
Once this point is reached, a \verb|StopIteration| exception is launched from within \verb|fun|.
This exception passes along the next set of positions.
Note that the positions are sent as \verb|additions| to the positions as a shift.
Since it is an addition to the existing object, it is required to set changes with either \verb|+=| or \verb|-=| operators.

\begin{lstlisting}[language=Python, caption={Illustration of linking scipy optimisers with VASP Python Plugins.}, label={lst:illustration_scipy}]

import numpy as np
from scipy.optimize import minimize

history = {}
x_start = None

def structure(constants, additions):
    global x_start
    x_curr = constants.positions @ constants.lattice_vectors
    history[tuple(x_curr)] = (constants.total_energy, -constants.forces)
    if x_start is None: x_start = x_curr

    def fun(x): 
        if tuple(x) in history: return history[tuple(x)][0]
        raise StopIteration(x)

    def jac(x): 
        return history[tuple(x)][1]

    try:
        minimize(fun, x0=x_curr, jac=jac, method="CG")
    except StopIteration as next_step:
        new_x = next_step.args[0]
        shift = (new_x - x_curr) @ np.linalg.inv(constants.lattice_vectors)
        additions.positions += shift
\end{lstlisting}

\begin{figure}[!htb]
    \centering
    \includegraphics[width=\linewidth]{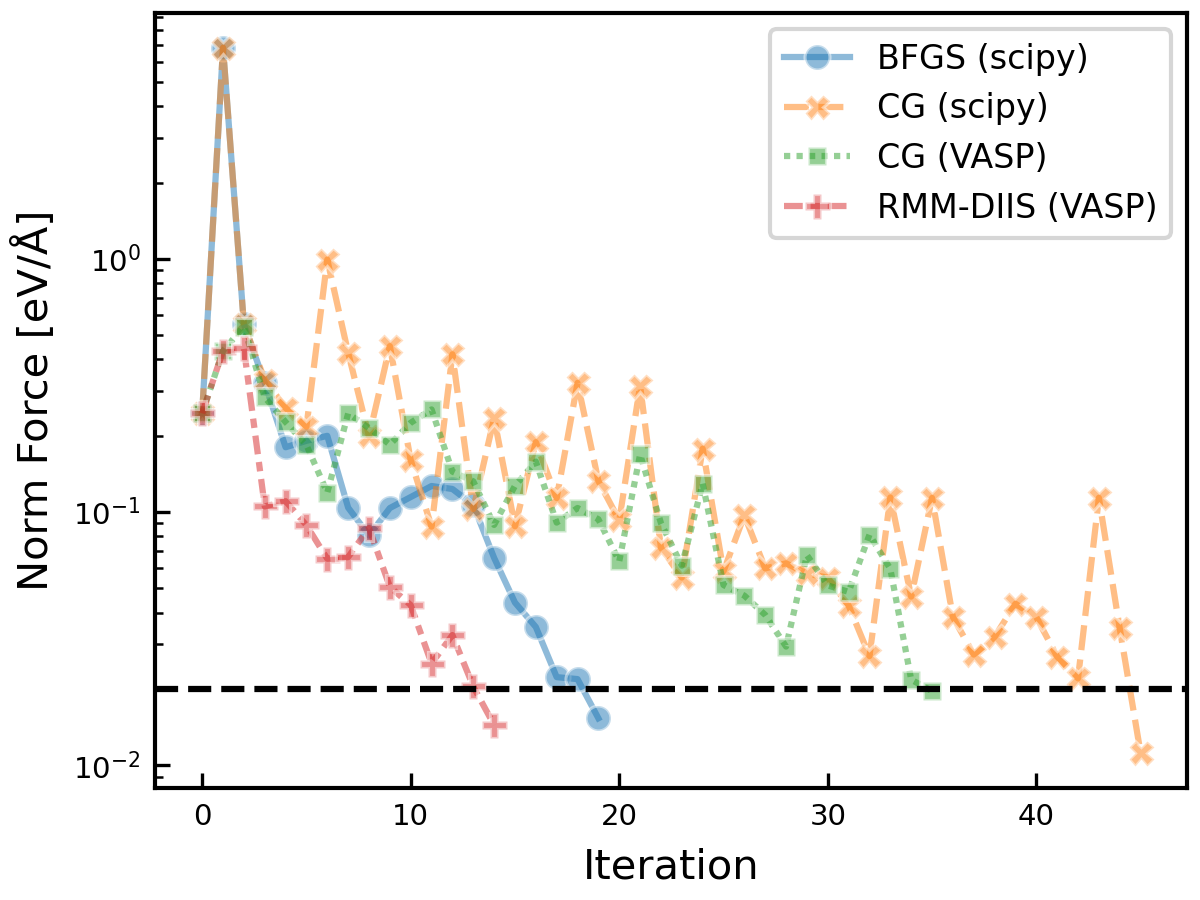}
    \caption{Comparison of the norm of the forces between existing optimisers in VASP and those linked with \texttt{scipy}.}
    \label{fig:scipy_test}
\end{figure}

Figure \ref{fig:scipy_test} shows the performance of four optimisers, RMM-DIIS implemented in VASP (red), CG implemented in VASP (green), CG implemented in \verb|scipy| (orange) and BFGS implemented in \verb|scipy| (blue).
For this test, we choose a structure from the Open Catalyst Dataset (index: random585454).
We find that the manually optimised implementations within VASP converge in fewer iterations (less than 20) as compared with generic implementations such as \verb|scipy| (37 or higher) for this particular example.

\subsection{Implicit solvent model using the \texttt{local\_potential} plugin}

\begin{figure*}[!htb]
    \centering
    \includegraphics[width=0.7\linewidth]{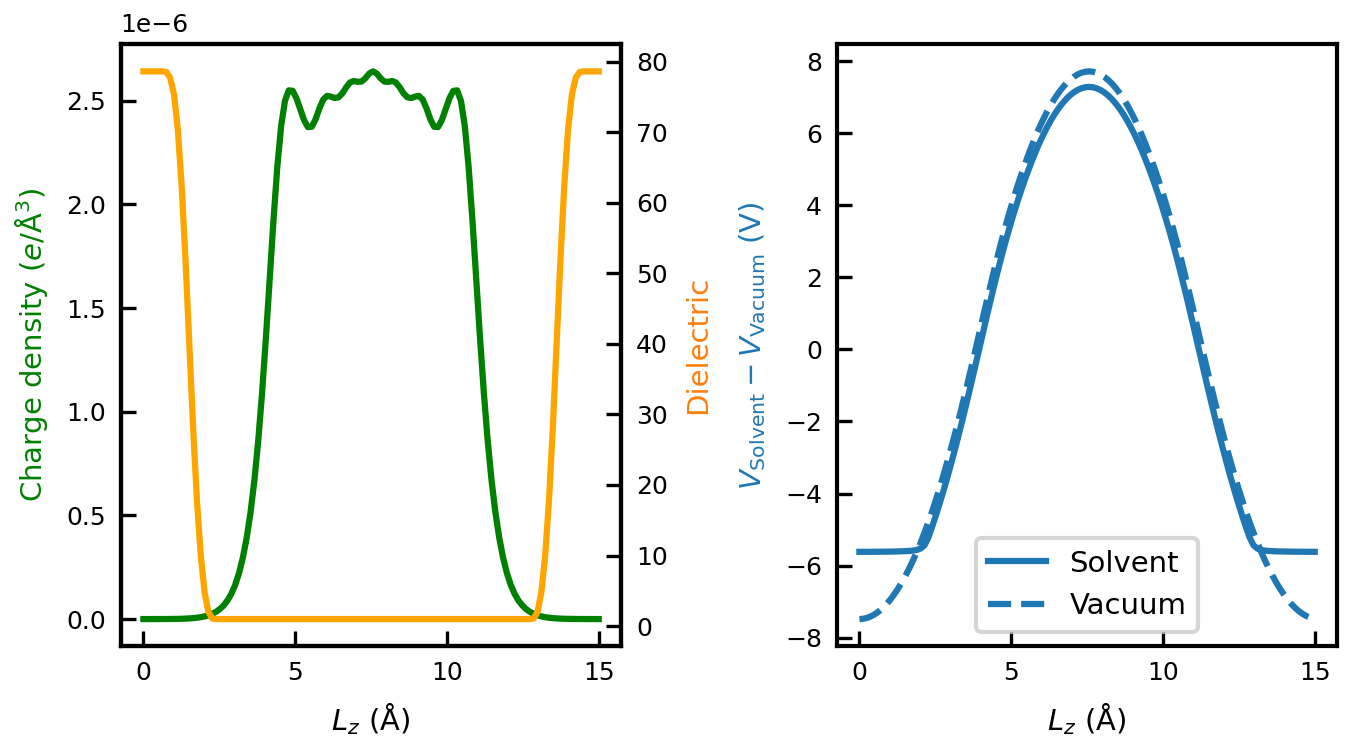}
    \caption{SCCS generated dielectric (orange), charge density (green) and computed vacuum (dashed blue) and solvent (solid blue) potential.}
    \label{fig:implicit}
\end{figure*}

In this illustration, we implement the self-consistent continuum solvent (SCCS) model\cite{Andreussi2012} and interface it with VASP.
Listing \ref{lst:illustration_implicit} shows a sketch of the code required to perform this link in practice.
First, a cavity is created based on the charge density using the algorithm in Ref \onlinecite{Andreussi2012}.
This cavity and the charge density is shown in Figure \ref{fig:implicit} for an Al(111) face-centered cubic slab at the center of the cell.
The dielectric (in orange) is close to the bulk dielectric in areas where the charge density (green) is approximately zero ($L_z<2\AA$ and $L_z>13\AA$).
Similarly, where the charge density is significant, the dielectric is close to 1 (indicating the absence of solvent).

This constructed dielectric and the charge density, is passed along to a generalised Poisson solver.
\verb|v_vac| (dashed blue, Figure \ref{fig:implicit}) stores the solution to the Poisson equation under vacuum conditions, i.e.\, the solvent dielectric is one.
\verb|v_solv| (solid blue, Figure \ref{fig:implicit}) stores the solution to the Poisson equation under solvent boundary conditions.
Note that the field (the gradient of the potential) flattens out for the solvent ($L_z<2\AA$ and $L_z>13\AA$).
This flattening is a consequence of the comparatively large bulk dielectric constant (78.6) used in this work to mimic water.

Once \verb|v_vac| and \verb|v_solv| are computed, their difference is sent back to VASP by adding it to the \verb|additions.total_potential| object.
Note that we subtract \verb|v_vac| from \verb|v_solv| as VASP already computes and stores \verb|v_vac|.
The subtraction ensures that \verb|v_vac| is removed and \verb|v_solv| takes its place. The resulting potential used by VASP will correspond to that generated in the presence of an implicit solvent.

\begin{lstlisting}[language=Python, caption={Illustration of linking implicit solvation models with VASP}, label={lst:illustration_implicit}]

def apply_solvation(constants, additions):
    lattice = constants.lattice_vectors
    rho = constants.charge_density
    sccs = SCCSDielectric(rho_min=1e-4, rho_max=1e-2, eps_bulk=78.6)
    eps_grid = sccs.get_dielectric_grid(rho)
    
    # A poisson solver with different boundary conditions
    solver = Poisson(lattice)
    v_solv = solver.solve(rho, dielectric_grid=eps_grid, solver="cg")
    v_vac = solver.solve(rho, boundary_condition="tin-foil-periodic")

    v_diff = (v_solv - v_vac)
    additions.total_potential += v_diff
\end{lstlisting}

\subsection{Dispersion corrections with the \texttt{force\_and\_stress} plugin}

Finally, we discuss how the \verb|force_and_stress| plugin interfaces with external pair-wise force modifiers.
For this illustration, we link the \verb|force_and_stress| plugin with the \verb|DFT-D4|\cite{caldeweyherGenerallyApplicableAtomiccharge2019,caldeweyherExtensionEvaluationD42020}

\begin{lstlisting}[language=Python, caption={Illustration of linking DFT-D4 with VASP}, label={lst:illustration_d4}]
from vasp.force_field import AseForceField

def force_and_stress(constants, additions):
    calculator = DFTD4(method="pbe")
    force_field = AseForceField(calculator)
    force_field.force_and_stress(constants, additions)
\end{lstlisting}

\begin{figure}[!htb]
    \centering
    \includegraphics[width=0.7\linewidth]{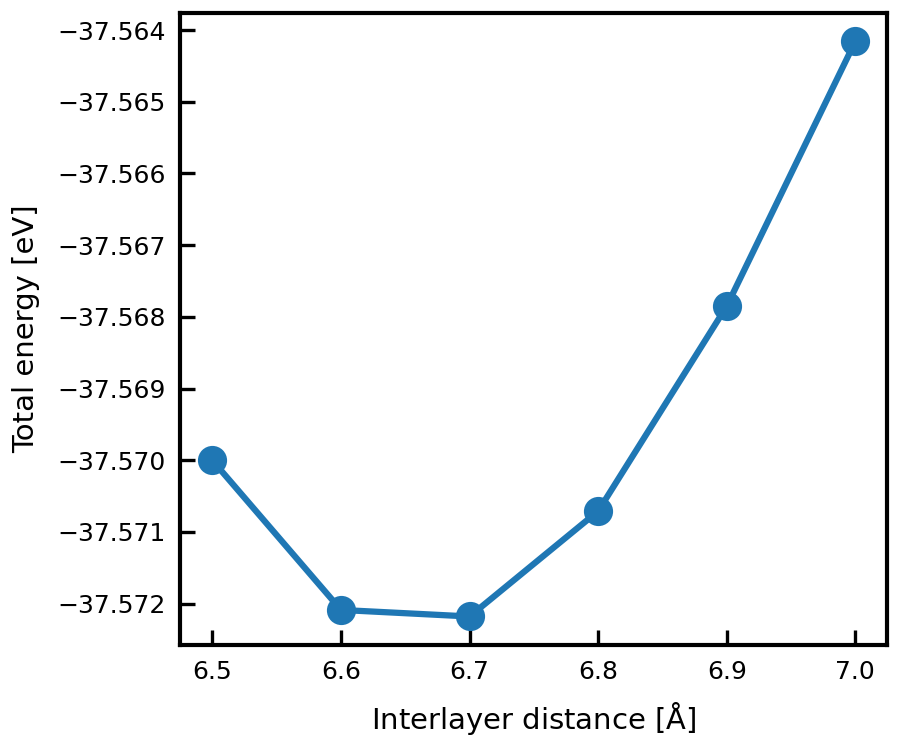}
    \caption{Change in the total energy with graphite interlayer spacing generated using the \texttt{force\_and\_stress} plugin.}
    \label{fig:illustration_d4}
\end{figure}

Listing \ref{lst:illustration_d4} shows the code required to perform a minimal link between \verb|DFT-D4| and VASP.
First, we import an \verb|AseForceField| from the VASP Plugin package.
This \verb|AseForceField| takes in any ASE calculator that can compute the forces and optionally, the stress tensor.
Finally, the force field adds the forces and the stress tensor generated by the calculator to those computed by VASP.

Figure \ref{fig:illustration_d4} shows the change in the total energy as a function of the interlayer distance between graphite.
This simple test is a commonly used benchmark to determine the accuracy of dispersion corrections such as \verb|DFT-D4|.
Using the PBE exchange correlation functional and \verb|DFT-D4| linked through the VASP Plugin infrastructure, we find that the interlayer distance is between $6.6$--$6.7$~\AA.

The diverse applications presented in this section demonstrate the extensibility of the VASP Plugin infrastructure across distinct stages the SCF and relaxation cycles.
Using the \verb|structure|, \verb|local_potential|, and \verb|force_and_stress| interfaces, we have shown how external Python libraries, ranging from generic implementation such as \verb|scipy| to specialised physical models like \verb|DFT-D4| and custom implicit solvation implementations can be integrated into existing VASP workflows. 
\section{Conclusions}

In this work, we develop a Python plugin infrastructure to link VASP, a DFT code written in Fortran, with Python functions.
This plugins infrastructure uses a C++ interface, enabled by the \verb|pybind11| library, to convert data structures in Fortran to those in Python.
Central to this implementation is the use of shared memory buffers and Python dataclasses, which provide a safe and efficient method for users to access unmodifiable constants and apply changes via additions.
We describe the flexibility of this approach by applying it to three common illustrations in electronic structure calculations.
First, we show that the \verb|structure| plugin allows for the  integration of external optimisation packages like \verb|scipy|.
Second, we illustrate that the \verb|local_potential| plugin enables the development of complex models, such as implicit solvation, entirely within the Python environment while remaining fully coupled to the VASP SCF cycle.
Finally, we show that the \verb|force_and_stress| plugin simplifies the inclusion of energy terms, such as \verb|DFT-D4| dispersion corrections.

By lowering the barrier to entry for source-code modification, this VASP Python plugin system allows researchers to implement custom algorithms without requiring deep expertise in legacy Fortran structure. 
Overall, we believe this framework provides a robust foundation for future developments in materials discovery, allowing the community to develop and expand the scientific Python ecosystem alongside established DFT routines.
\section{Computational Methods}

All density functional theory calculations were performed with the Vienna \textit{ab-initio} Simulation Package (VASP).\cite{Kresse1996}
The Python plugins feature described in this manuscript is implemented in VASP and is available as of version 6.5.0.

Relaxation calculations to illustrate the differences between ionic solvers were performed a plane-wave cutoff of 520~eV.
For this illustration we use a cell with stoichiometry Zr$_{24}$Sc$_8$ with an NHO molecule placed as an adsorbate on a top site.
The energy convergence threshold for each SCF cycle was set to $10^{-6}$~eV.
The \verb|Zr_sv|, \verb|Sc_sv|, \verb|H|, \verb|N| and \verb|O| pseudopotentials were used for Zr, Sc, H, N and O respectively.
An Al(111) slab was used to illustrate the SCCS solvent implementation.
A plane-wave cutoff of 241~eV was used. The \verb|Al| pseudopotential was used for this calculation.
To compute the graphite interlayer distance, we use the \verb|C| pseudopotential
with its default cutoff of 400~eV and a $16 \times 16 \times 8$ $\Gamma$-centered k-point mesh. We converge the energy to $10^{-6}$~eV with \verb|PREC = Accurate|.
The PBE\cite{Perdew1996} functional was used for all calculations.
Gaussian smearing was used with a broadening of 0.1~eV for all calculations except for the graphite interlayer distance where the width was reduced to 0.01~eV.

\begin{acknowledgments}
We acknowledge support form EuroHPC Joint Undertaking for awarding us access to Leonardo at CINECA, Italy.
S.V. acknowledges funding from ANRF MATRICS (ANRF/ARGM/2025/000139/TS) and seed grant (RD/0524-IRCCSH0-023) at IIT Bombay.
\end{acknowledgments}

\section*{Author contribution}

\textbf{Sudarshan Vijay}: Conceptualization, Data curation, Formal analysis, Investigation, Software, Methodology, Validation, Visualization, Writing – original draft, Writing – review \& editing. 
\textbf{Martijn Marsman}: Formal analysis, Software, Writing – review \& editing.
\textbf{Georg Kresse}: Conceptualization, Funding acquisition,  Methodology, Resources
\textbf{Martin Schlipf}: Conceptualization, Data curation, Formal analysis, Investigation, Software, Methodology, Validation, Visualization, Writing – original draft, Writing – review \& editing. 

\section*{Use of AI Tools}
We used AI tools for grammar and language checking in the preparation of this manuscript.
We have reviewed and edited the content as needed and take full responsibility for the 
content of this publication.

\bibliography{references_utf}

\end{document}